\documentclass[12pt]{article}
\let\section=\subsection  \let\subsection=\subsubsection    
\newcommand{\beq}{\begin{equation}}
\newcommand{\eeq}[1]{\label{#1} \end{equation}}

\newcommand{\insertplotllargee}[1]{
  \centerline{\psfig{figure={#1},height=18cm}}
}
\newcommand{\insertplotwide}[1]{
  \centerline{\psfig{figure={#1},width=14cm}}
}

\newcommand{\dlt}{\bigtriangleup}

\usepackage{graphicx,psfig,epsf}

\begin{document}
\begin{center}
 {\large \bf Initial conditions for RHIC collisions\footnote{
 To appear in the Proceedings of the 
 30th International Workshop on Gross Properties of Nuclei and 
 Nuclear Excitation: Hirschegg 2002: 
 "Ultrarelativistic Heavy Ion Collisions", Hirschegg,
 Austria, 13-19 Jan 2002.},}\\[5mm]
 V.K.~MAGAS$^{1,2}$, L.~P.~CSERNAI$^{2,3}$ and D.~D.~STROTTMAN$^4$\\[5mm]
 {\small \it $^1$ Center for Physics of Fundamental Interactions
(CFIF), Physics Department \\
 Instituto Superior Tecnico, Av. Rovisco Pais, 
1049-001 Lisbon, Portugal \\[2mm] }
 {\small \it $^2$ Section for Theoretical and Computational Physics,
 Department of Physics\\
 University of Bergen, Allegaten 55, N-5007, Norway\\[2mm] }
 {\small \it $^3$ KFKI Research Institute for Particle and Nuclear
Physics\\
 P.O.Box 49, 1525 Budapest, Hungary\\[2mm] }
 {\small \it $^4$ Theoretical Division, Los Alamos National Laboratory\\
 Los Alamos, NM, 87454, USA\\[8mm] }
\end{center}

\begin{abstract}\noindent
An Effective String Rope Model (ESRM) for heavy 
ion collisions at RHIC energies is
presented. Our model takes into account baryon recoil for both target and
projectile, arising from the acceleration of partons in an effective field,
produced in the collision. The typical field strength (string
tension) for RHIC energies is about $5-12 \ GeV/fm$, what allows us to talk
about ``string ropes''. Now we describe an "expanding final streaks" scenario
\cite{MCS02}, in contrast to a "non-expanding final streaks" discussed in
Ref. \cite{MCS01}. The results show that a QGP forms a tilted disk, such
that the direction of the largest pressure gradient stays in the reaction
plane, but deviates from both the beam and the usual transverse flow
directions. The produced initial state can be used as an initial condition
for further hydrodynamical calculations. Such initial conditions lead to the
creation of a third flow component.
\end{abstract}

The realistic and detailed description of an energetic heavy ion reaction
requires a Multi Module Model, where the different stages of the reaction are
each described with a suitable theoretical approach. It is important that
these Modules are coupled to each other correctly: on the interface, which is
a three dimensional hypersurface in space-time with normal $d\sigma^\mu$, all
conservation laws should be satisfied, and entropy should not decrease. These
matching conditions were worked out and studied for the matching at FO
hypersurfaces in details in Refs.
\cite{FO1,FO3}. 

In energetic collisions of large heavy ions, one-fluid dynamics is a valid and
good description for the intermediate stages of the reaction. Here,
interactions are strong and frequent, so that other models, (e.g. transport
models, string models, etc., that assume binary collisions, with free
propagation of constituents between collisions) have limited validity. On the
other hand, the initial and final, Freeze Out (FO), stages of the reaction
are outside the domain of applicability of the fluid dynamical model. For the
highest energies achieved nowadays at RHIC, hydrodynamic calculations give a
good description of the observed radial and elliptic flows 
\cite{Sollfrank-BigHydro,Schlei-BigHydro,Kolb-Radial,Htoh}. 

The initial stages for RHIC energies are the most problematic. Non of the
theoretical models currently on the physics market can unambiguously describe
the initial stages (see \cite{MCS02} for  a discussion). 

\section{Expanding final streaks in ESRM}
In Refs. \cite{MCS00,CAM00,MCS01} we assumed that the final result of the
collision of two streaks, after stopping the string's expansion and after its
decay, is one streak of the length $\dlt l_f$ with homogeneous energy density
distribution, $e_f$, moving like one object with rapidity
$y_f$ ($e_f$ and $y_f$ can be easily found from conservation laws).  The
typical values of the string tension, $\sigma$, are of the order of $10\
GeV/fm$, and these may be treated as several parallel strings.  We assumed
that such a final state is due to string-string interactions  and string
decays, which we are not going to describe in our simple model. One of the
simplest ways to quantitatively take into account string  decays is presented
in Ref. \cite{mishkap}.

Now we would like to describe the scenario with expanding final streaks, 
which seems to be more realistic \cite{MCS02}. It is based on the solution
for the one-dimensional expansion of the finite streak into the vacuum, 
which can be found \cite{MCS02} generalizing the description in \cite{R95}.

The initial condition is
\begin{eqnarray} \label{in1}
e (z,t_0) & = & \left\{ \begin{array}{ll}
  0&~~-\infty <z < R_-~,\\
		 e_0 &,~~R_-\leq z \leq R_+ \\
  0 &,~~R_+ <z < \infty~,
  \end{array} \right. \\
v(z,t_0) & = & \left\{ \begin{array}{ll}
  -1 &,~~-\infty < z< R_- \\
  \tanh y_0 &,~~R_- \leq z \leq R_+ \\
  1 &,~~R_+ <z < \infty~,
  \end{array} \right. \label{in2}\ ,
\end{eqnarray}
where $R_-$, $R_+$ are the borders of the system.

For the forward-going rarefaction wave, $v>0$, generated on the right end of
the initial streak, we infer that the head of this wave (the point where the
rarefaction of matter starts, i.e., where the energy density $e$ starts to
fall below $e_0$) travels with the velocity $(\tanh y_0-c_0)/(1-c_0\tanh
y_0)$ to the left. On the other hand, the base of the rarefaction wave (the
point where $e$ starts to acquire non-vanishing values) travels with light
velocity $v = 1$ to the right. The solution for a simple wave has a 
similarity form \cite{R95}, i.e., the profile of the rarefaction wave does not
change with time when plotted as a function of the similarity variable
$\zeta_\pm \equiv
\frac{z\mp-R_\pm}{t-t_0}$. Thus, 
\begin{equation}
e(\zeta_+) = e_0 \cdot \left\{ \begin{array}{ll}
  1 &,~~-1 \leq \zeta_+ \leq \frac{\tanh y_0-c_0}{1-c_0\tanh y_0} \\
 {\displaystyle \left[\frac{1+\tanh y_0}{1-\tanh y_0}~ \frac{1-c_0}{1+c_0}~
	 \frac{1-\zeta_+}{1+\zeta_+}
  \right]}^{(1+c_0^2)/2c_0} &,~~
		 \frac{\tanh y_0-c_0}{1-c_0\tanh y_0}< \zeta_+ \leq 1~.
  \end{array} \right.
\label{epl}
\end{equation}
\beq
\tanh y(\zeta_+) = \left\{ \begin{array}{ll}
  \tanh y_0 &,~~-1 \leq \zeta_+ \leq \frac{\tanh 
y_0-c_0}{1-c_0\tanh y_0} \\
 {\displaystyle \frac{c_0+\zeta_+}{1+\zeta_+ c_0}}&,~~
		 \frac{\tanh y_0-c_0}{1-c_0\tanh y_0}< \zeta_+ \leq 1~.
  \end{array} \right.
\eeq{ypl}

Similarly for the backward-going rarefaction wave, $v<0$, we get
\begin{equation}
e(\zeta_-) = e_0 \cdot \left\{ \begin{array}{ll}
  1 &,~~\frac{\tanh y_0+c_0}{1+c_0\tanh y_0} \leq \zeta_- \leq 1 \\
 {\displaystyle \left[\frac{1-\tanh y_0}{1+\tanh y_0}~ \frac{1-c_0}{1+c_0}~
	 \frac{1+\zeta_-}{1-\zeta_-}
  \right]}^{(1+c_0^2)/2c_0} &,~~
		 -1< \zeta_- \leq \frac{\tanh y_0+c_0}{1+c_0\tanh y_0}~,
  \end{array} \right.
\label{emin}
\end{equation}
\beq
\tanh y(\zeta_-) = \left\{ \begin{array}{ll}
  \tanh y_0 &,~~ \frac{\tanh y_0+c_0}{1+c_0\tanh y_0} \leq 
\zeta_- \leq 1\\
 {\displaystyle \frac{\zeta_--c_0}{1-\zeta_- c_0}}&,~~
		 -1< \zeta_- \leq \frac{\tanh y_0+c_0}{1+c_0\tanh y_0}~.
  \end{array} \right.
\eeq{ymin}

This simple analytical solution is valid as long as the two rarefaction waves
did not overlap in the middle of the system. Further evolution becomes more
complicated and doesn't have a similarity form any more.

\begin{figure}[htb]
\insertplotwide{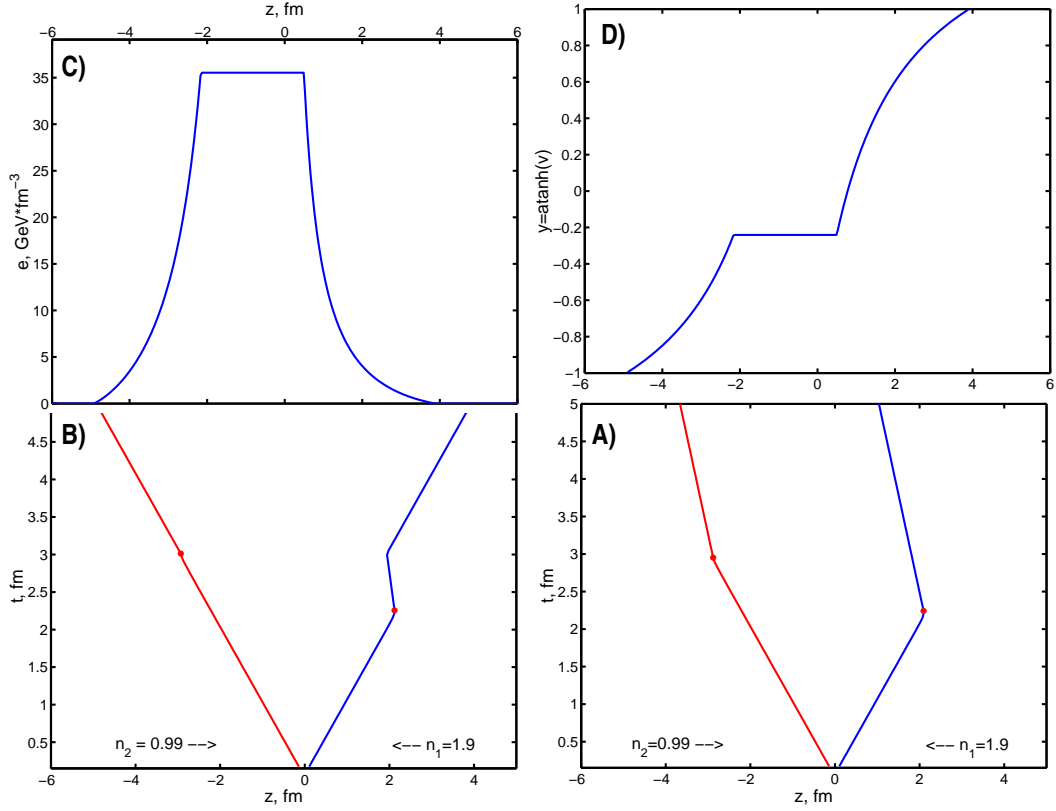}
\caption[]{A) The typical trajectory of the ends of two initial streaks
corresponding to the numbers of nucleons $n_1$ and $n_2$,
$\varepsilon_0=65\ GeV/nucl$, $A=0.09$ (see ESRM description, for example
\cite{MCS01}). Stars denote the points where $y_{1,2}=y_f$. From
$t=t_0$ until these stars, the streak ends move according trajectories
derived in Ref. \cite{MCS01} (see eqs. (37, 38)). Then the final streak
starts to move like a single object with rapidity $y_f$
in CRF. B) The same situation as in subplot (A), but for expanding final
streak. C) Shows $e(z)$ profiles of expanding final streak 
($t_h=5 \ fm$). We can clearly
see three regions - two of forward and backward rarefaction waves and a
middle where the initial energy density, $e_f$, is still preserved.
D) Shows the rapidity profile of the expanding final streak ($t_h=5 \ fm$). 
We can clearly
see three regions - two of forward and backward rarefaction waves and a
middle where the initial rapidity,
$y_f$, is still preserved. }
\label{fig2}
\end{figure}

Thus, based on the solution presented above we can have a more advanced
description of the final streaks.  Let us assume that the homogeneous final
streaks, with some $e_f$, $y_f$, are formed in the Center-of-Rapidity frame
(CRF)  when the larger of initial streaks reach the rapidity $y_f$. Up to
this point the fluid cell trajectories are the same as those for
non-expanding scenario \cite{MCS01}, but after the homogeneous final streak is
formed, it starts to expand according to the simple rarefaction wave
solution,  eqs. (\ref{epl}-\ref{ymin}) \cite{MCS02}. Figure \ref{fig2} (B)
shows new trajectories of the streak ends  (compare to (A) from Ref.
\cite{MCS01}), and Figures \ref{fig2} (C,D) present energy density and
rapidity profile for this expanding streak.

Such an initial state with expanding streaks will also help to avoid the
problem which may cause the development of numerical artifacts, namely a
step-like in the beam direction initial energy density distribution (output of
ESRM): it has a jump from $e_f=const$ inside the matter, to $0$ in the outside
vacuum (of course, where there is also a jump of $e$ as a 
function of $x,y$, but it
is much smoother and this is not the direction of the initial expansion). In
order to avoid (or at least to suppress) this effect, it was proposed
in Ref. \cite{hirano} to smooth over initial energy density distribution, for
example by a Gaussian shape. Our simple analytic solution smoothes out this
jump in a natural way.

\begin{figure}[htb]
\insertplotllargee{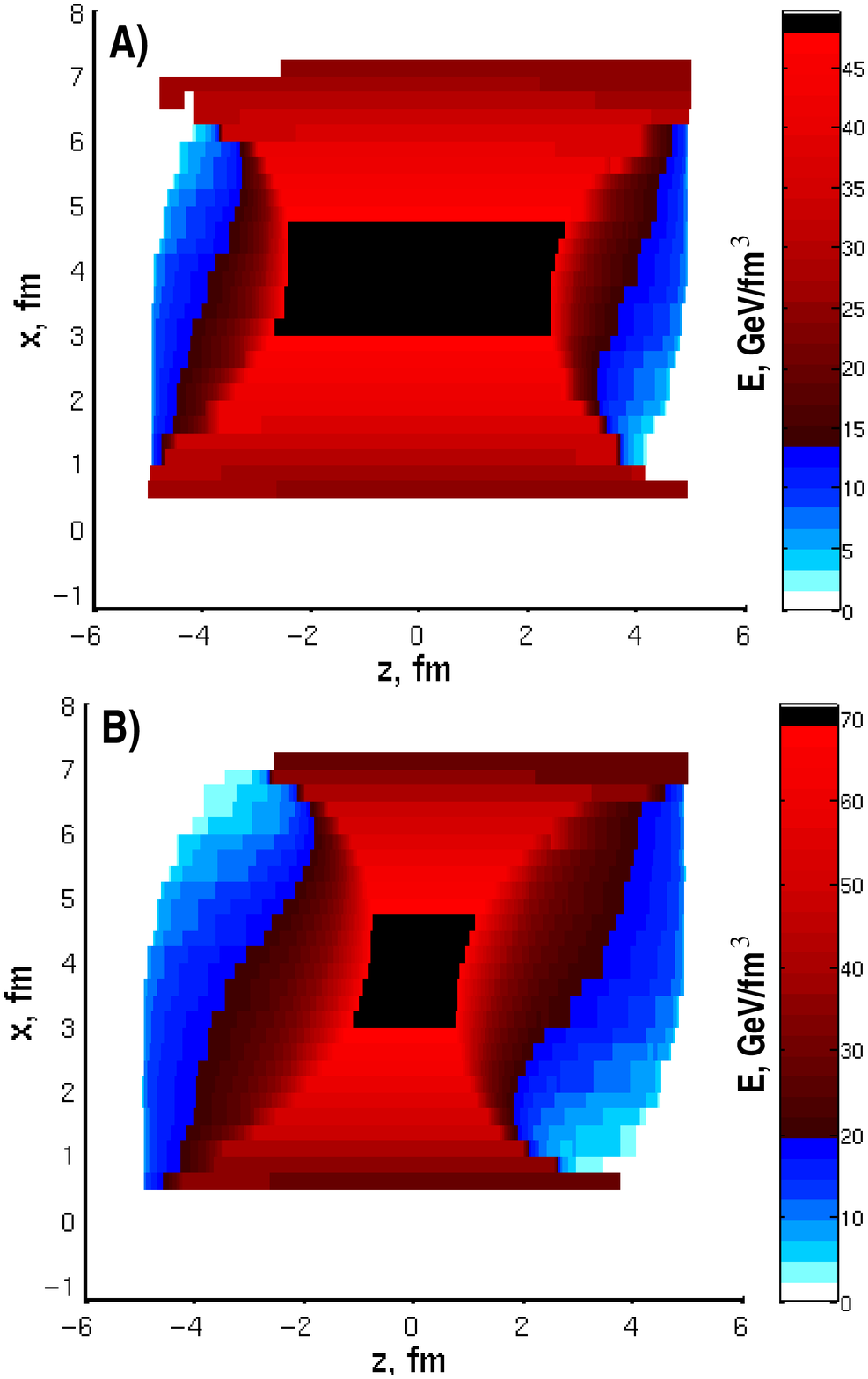}
\vspace{-0.5cm}
\caption[]{Au+Au collision at $\varepsilon_0=100\ GeV/nucl$,
$\left(b=0.5\cdot2\ R_{Au}\right)$ ,
$E=T^{00}$ is presented in the reaction plane
as a function of $x$ and $z$ for $t_h=5\ fm/c$.
Subplot A) $A=0.065$,
subplot B) $A=0.08$. The QGP volume has a
shape of a tilted disk
and may produce a third flow component. }
\label{res1}
\end{figure}

Fig. \ref{res1} shows the energy density distribution in the reaction plane
for the RHIC collisions  calculated in the scenario with expanding final
streaks.  The QGP forms a tilted disk for $b\not =0$. Thus, the direction of
fastest expansion, the same as the largest pressure gradient, will be in the
reaction plane, but will deviate from both the beam axis and the usual
transverse flow direction. So, the new ``third flow component'' 
\cite{CR99} appears in addition to the usual transverse flow component in the
reaction plane. For non-expanding streaks this was shown in \cite{CAMS02}. If
we let final streaks expand, this smoothes out the picture, but most of the
matter, nevertheless, keeps a similar energy density profile and velocity
distribution.

\end{document}